\documentclass[prb, twocolumn, superscriptaddress, showpacs]{revtex4}  
\usepackage{amssymb}
\usepackage[pdftex,colorlinks,bookmarks=false,citecolor=blue,linkcolor=red,urlcolor=blue]{hyperref}
\usepackage{epsfig}
\usepackage{epstopdf}
\usepackage{times}
\usepackage{amsmath}
\usepackage{bbm}
\usepackage{graphicx}
\usepackage{color}
\usepackage{enumerate}

\begin{document}

\author{Philipp Werner}
\affiliation{Theoretische Physik, ETH Zurich, 8093 Z{\"u}rich, Switzerland}
\author{Takashi Oka}
\affiliation{Department of Physics, Tokyo University, Hongo, Tokyo 113-0033, Japan}
\author{Martin Eckstein}
\affiliation{Theoretische Physik, ETH Zurich, 8093 Z{\"u}rich, Switzerland}
\author {Andrew J. Millis}
\affiliation{Department of Physics, Columbia University, 538 West, 120th Street, New York, NY 10027, USA}

\title{Weak-coupling quantum Monte Carlo calculations on the Keldysh contour: theory and application to the current-voltage characteristics of the Anderson model}

\date{\today}

\hyphenation{}

\begin{abstract}
We present optimized implementations  of the weak-coupling continuous-time Monte Carlo method defined for nonequilibrium problems on the Keldysh contour. We describe and compare two methods of preparing the system before beginning the real-time calculation: the  ``interaction quench"  and the ``voltage quench", which are  found to be suitable for large and small voltage biasses, respectively. We also discuss technical optimizations which increase the efficiency of the real-time measurements. The methods  allow the accurate simulation of transport through  quantum dots over  wider interaction ranges and longer times than have heretofore been possible. The current-voltage characteristics of the particle-hole symmetric Anderson impurity model is presented for interactions $U$ up to $10$ times the intrinsic level width $\Gamma$. We compare the Monte Carlo results to fourth order perturbation theory, finding that perturbation theory begins to fail at  $U/\Gamma\gtrsim 4$. Within the parameter range studied we find no evidence for a splitting of the Kondo resonance due to the applied voltage. The interplay of voltage and temperature and the Coulomb blockade conductance regime are studied.
\end{abstract}


\maketitle

\section{Introduction}

The development of robust methods for the computation of nonequilibrium properties of quantum many-particle systems is a crucial issue in present-day condensed matter physics, with impact on topics ranging from nonequilibrium transport in nanostructures\cite{Goldhaber-Gordon98}  to pump-probe spectroscopy of  bulk condensed matter systems\cite{Iwai03, Iwai06} and the wide range of new spectroscopies possible in cold atom systems.\cite{Schneider08}  An important step forward occurred with the development of continuous-time quantum Monte Carlo (CTQMC) methods for impurity models. These algorithms were first introduced as imaginary-time methods for obtaining  equilibrium properties \cite{Rombouts99,Rubtsov05,Werner06,Gull08_ctaux} and soon afterwards were extended to real-time dynamics and  nonequilibrium problems.\cite{Muehlbacher08,Schmidt08,Werner09, Schiro09} 

The continuous-time methods are in essence  stochastic samplings of diagrammatic expansions  of the time evolution operator. The mean perturbation order required in the calculation increases with the time (or inverse temperature) to be studied and the calculations are limited by the perturbation order which can be achieved with given computational resources.   In the equilibrium case one considers the imaginary-time evolution operator $\exp[-\tau H]$ which is real and positive definite, so the computational task  is to estimate a sum of real (decaying) exponentials and the only sign problem which arises is the fermion sign problem occurring in models complicated enough to sustain fermion loops. For these reasons the CTQMC methods have proven to be very powerful in the equilibrium context.\cite{Gull07_comparison}   In the real time case, on the other hand, one must consider the intrinsically complex time evolution operator $\exp[-itH]$, and convergence comes from the cancellation of oscillations. The theoretical task is therefore to estimate the sum of terms with oscillating signs or rotating phases and a severe ``dynamical" sign problem occurs even in the absence of fermion loops. The average sign decreases exponentially with perturbation order, which limits the accessible range of interaction strengths and simulation times. 

Because of these limitations, important questions such as the nonequilibrium Kondo effect could so far not be adequately addressed. The equilibrium Kondo effect in quantum dots,\cite{Ng88, Glazman88} which involves the formation of a scattering resonance (density of states peak) at the Fermi level, was experimentally confirmed in the zero bias limit.\cite{Goldhaber-Gordon98} While at very low voltage biasses, the pinning of the Kondo resonance to the Fermi level leads to an unrenormalized conductance for symmetric dots, it is well known that a high voltage bias destroys the Kondo effect. The crossover from the low voltage universal regime (``linear response regime") to the higher bias Coulomb blockade regime is presently not understood. It has been proposed on the basis of the noncrossing approximation,\cite{Meir93} real time diagrammatic methods,\cite{Konig96} and perturbative
calculations\cite{Fujii03} that the peak in the density of states splits into two in a certain parameter regime. Our previous investigation of the non-equilibrium Anderson model \cite{Werner09} produced no sign of this phenomenon, but the accuracy of the simulations in the low-bias region was not sufficient to settle the issue. Methodological improvements allowing a more accurate  numerical study of the small-to-intermediate voltage regime are therefore needed. 

The existing continuous-time Quantum Monte Carlo approaches for nonequilibrium systems are more-or-less direct extensions of the imaginary time algorithms previously developed. It appears worthwhile to attempt to optimize them, even though  the dynamical sign problem inherent in these methods unavoidably limits what can be achieved. In this paper we present an efficient implementation of the weak-coupling diagrammatic Monte Carlo method for non-equilibrium systems, describing ways to reorganize the expansion and to improve the measurement formulae in order to increase the accuracy of the numerical data for a given set of parameters.   

The method introduced previously\cite{Werner09} corresponds to the simulation of a system prepared in the nonequilibrium but non-interacting state, with the interaction turned on at time $t=0$. We refer to this simulation method as an ``interaction quench". Since the real-time methods compute the time evolution of the system after the quench, an important consideration is the time needed for the system to evolve to the interacting steady state.  Optimized preparation of the initial ensemble has the potential to reduce this relaxation time, therefore leading to  simulations requiring a smaller total time interval for the measurement of a given property.  Motivated by this idea we extend the formalism from two real-time branches to an ``L-shaped" contour which includes an imaginary time branch. Evolution along the imaginary time branch may be thought of as preparing the system in a correlated equilibrium state,  after which the voltage is turned on at time $t=0$. We refer to this simulation method as a ``voltage quench".  

One purpose of this paper is to compare interaction and voltage quenches. We will show that at temperature $T=0$ interaction quenches are suitable for voltage biasses larger than the Kondo temperature (i.~e. for voltage biasses large enough to suppress the Kondo resonance in the many-body density of states). The times which can be reached in the Monte Carlo simulation are long enough to observe convergence into a steady state even at large interaction strengths. On the other hand, if  the voltage bias is small and the temperature is finite, the voltage quench is a suitable alternative, because it allows the important ground state correlations to be built up via the computationally less problematic  imaginary time evolution.  

We show that our  optimized implementation allows the  computation of  accurate currents over a wide voltage range, even for interaction strengths which are clearly outside the reach of low-order perturbation theory. We use the numerical results to test predictions based on fourth order perturbation theory in the self-energy.\cite{Fujii03} We determine the largest interaction strength for which the perturbation theory provides accurate results over the entire voltage range, and for larger interactions, the voltage window where deviations appear. The predicted splitting of the Kondo resonance \cite{Meir93,Konig96,Fujii03} is not evident in the numerical data.  

The rest of this paper is organized as follows. In section \ref{Model} we introduce the model to be solved and present the methods used to solve it, in particular defining the voltage and interaction quenches. Sections \ref{interactionquench} and \ref{voltagequench}   present results for the interaction and voltage quenches respectively. Section \ref{iv} gives results for the current-voltage characteristics of the model and section \ref{conclusions} is a summary and conclusion.

\section{Model and methods\label{Model}}

\subsection{Model}

We consider the one-orbital Anderson impurity model, which describes a single spin-degenerate ($\sigma$) level with a Hubbard interaction $U$ (the ``dot")  coupled by hybridization $V$  to two reservoirs (``leads") labeled by $\alpha=L,R$. The Hamiltonian $H_{QI}=H^0_\text{dot}+H_U+H_\text{bath}+H_\text{mix}$ of this model contains the terms
\begin{eqnarray}
H_\text{bath} &=& \sum_{\alpha=L,R} \sum_{p,\sigma} \big(\epsilon^\alpha_{p,\sigma}-\mu_\alpha \big)a^{\alpha \dagger}_{p,\sigma} a^\alpha_{p,\sigma},\label{H_bath}\\
H_\text{mix} &=& \sum_{\alpha=L,R} \sum_{p,\sigma} \big(V_p^\alpha a^{\alpha \dagger}_{p,\sigma}d_\sigma+h.c. \big),\label{H_mix}\\
H^0_\text{dot}&=&\epsilon_d\sum_\sigma n_{d,\sigma},\label{H_d}\\
H_{U} &=& U(n_{d,\uparrow} n_{d,\downarrow}-(n_{d,\uparrow}+n_{d,\downarrow})/2).\label{H_U}
\end{eqnarray}
In the following we will consider two sources of time dependence in $H_{QI}$. In the interaction quench we take $U=0$ for times $t<0$ with an instantaneous step to a non-zero $U$ at $t=0$; in the voltage quench we take $\mu_L=\mu_R$ for time $t<0$ with an instantaneous step to a nonzero $\mu_L-\mu_R$ at $t=0$. We assume that the lead electrons equilibrate instantly to the new chemical potential so that the  equal time correlators of lead operators are $\langle a^{\alpha\dagger}_{p,\sigma}a^{\beta}_{p',\sigma'}\rangle=\delta_{\alpha,\beta}\delta_{p,p'}\delta_{\sigma,\sigma^{'}}f_{T_\alpha}(\epsilon^\alpha_{p,\sigma}-\mu_\alpha)$, with $f_T(x)=(e^{x/T}+1)^{-1}$ the Fermi distribution function for temperature $T$ and $\mu_\alpha$ the value of the chemical potential for lead $\alpha$ at the appropriate time. 

In this paper we will consider only symmetric voltage biasses ($\mu_L=-\mu_R=V/2$) and half-filled dots ($\epsilon_d=0$). The consequences of relaxing these assumptions will be briefly mentioned in the conclusions. 
The energy scales of the model are set by the level broadenings 
\begin{equation}
\Gamma^\alpha=\pi\sum_p|V_p^\alpha|^2\delta(\omega-\epsilon_p^\alpha)
\label{Gamdef}
\end{equation}
associated with the leads $\alpha$. The total level broadening 
\begin{equation}
\Gamma=\Gamma^L+\Gamma^R 
\label{Gammatotal}
\end{equation}
is used as the energy unit throughout the paper. 

We consider flat bands centered at zero, with a high energy cutoff $\omega_c$.  As we shall see in Section \ref{voltagequench} a sharp high frequency cutoff leads to oscillations in the time evolution of the current after a voltage quench. A sufficiently smooth band cutoff damps the oscillations but does not affect the steady state current. We adopt a Fermi-function like  smoothing with ``smoothing parameter" $\nu$, 
\begin{equation}
\Gamma^{L,R}(\omega)=\frac{\Gamma^{L,R}}{(1+e^{\nu(\omega-\omega_c)})(1+e^{-\nu(\omega+\omega_c)})}.
\end{equation}

\subsection{Real-time Monte Carlo method: weak-coupling approach}

We use the weak-coupling formulation of the real-time diagrammatic Monte Carlo approach as described in Ref.~\onlinecite{Werner09}. This is a real-time implementation of the continuous-time auxiliary field algorithm \cite{Gull08_ctaux} which is based on the combination of a weak-coupling expansion and an auxiliary field decomposition. Here, we will briefly summarize the main aspects of this method and then discuss some relevant issues concerning its efficient implementation. In order to enable simulations starting from an interacting initial state, we  formulate the method on the L-shaped contour which runs from $0$ to $t$ and back to $0$ along the Keldysh real time axis, then to $-i\beta$ along the imaginary time axis.  

\begin{figure}[t]
\begin{center}
\includegraphics[angle=0, width=0.8\columnwidth]{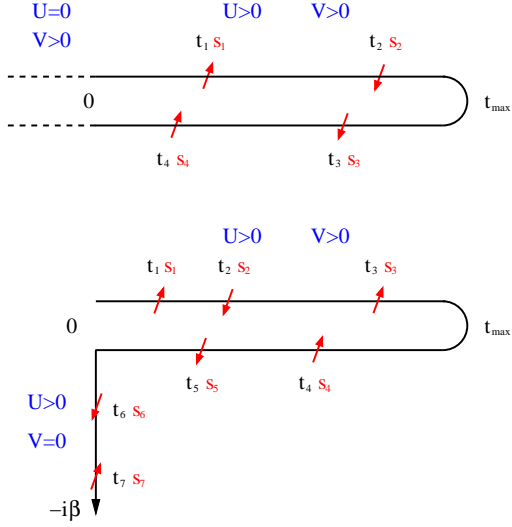}
\caption{Illustration of the Keldysh contour for the interaction quench (top panel) and voltage quench (bottom panel). In an interaction quench starting from $U=0$, the imaginary time branch of the contour is shifted to $t=-\infty$ and need not be explicitly considered in the Monte Carlo simulation. The red arrows represent auxiliary Ising spin variables. The top panel shows a Monte Carlo configuration corresponding to perturbation order $n_+=2$, $n_-=2$, and the bottom panel a configuration corresponding to $n_+=3$, $n_-=2$, $n_\beta=2$.}
\label{Lcontour}
\end{center}
\end{figure}

The weak coupling algorithm may be taken to  start from the following expression for the partition function $Z=Tr e^{-\beta H}$:
\begin{eqnarray}
Z &=& e^{K_\beta}Tr  \big[ e^{-\beta(H_\text{bath}^\text{eq}+H^0_\text{dot}+H_\text{mix}+H_{\tilde U}-K_\beta/\beta)} 
\nonumber \\
&&\hspace{12mm}\times e^{it(H_\text{bath}^\text{neq}+H^0_\text{dot}+H_U+H_\text{mix}-K_t/t)} \nonumber\\
&&\hspace{12mm}\times e^{-it(H_\text{bath}^\text{neq}+H^0_\text{dot}+H_U+H_\text{mix}-K_t/t)}\big],
\label{Z}
\end{eqnarray}
with $K_\beta$ and $K_t$ some arbitrary (non-zero) constants. 
The interaction and the  chemical potentials need not be the same on the imaginary time branch as they are on the real-time branches. In the formalism as written the interaction and chemical potentials are taken to  be time independent on the real-time branches, but it is straightforward to generalize the method to time-dependent $U$ and $\mu$. The notation $H_\text{bath}^\text{neq}$ indicates that on the real-time portion of the contour the two leads have different chemical potentials $\mu_\alpha=\mu_0\pm \delta \mu$, whereas $H_\text{bath}^\text{eq}$ means that on the imaginary time portion of the contour the two leads have the same chemical potential $\mu_0$. Henceforth we choose energies such that $\mu_0=0$ and consider a symmetrically applied bias voltage $V$ ($\delta \mu=V/2$). 

The time evolution along the real-time and imaginary-time contours is expanded in powers of $H_U-K_t/t$ and $H_U-K_\beta/\beta$, respectively. Each interaction vertex is then decoupled using Ising spin variables according to the formula\cite{Rombouts99} ($x=t$ or $\beta$)
\begin{eqnarray}
H_U-K_x/x &=& -\frac{K_x}{2x}\sum_{s=-1,1}e^{\gamma s (n_{d,\uparrow}-n_{d,\downarrow})},\\
\cosh(\gamma_x)&=&1+(xU)/(2K_x).
\label{decouple}
\end{eqnarray}
The resulting collection of Ising spin variables on the contour represents the Monte Carlo configuration $\{ (t_{1}, s_1),(t_{2}, s_2), \ldots (t_{n}, s_{n}) \}$, with $t_i$ denoting the position of spin $i$ on the L-shaped contour (see illustration in Fig.~\ref{Lcontour}). There are $n_+$ spins on the forward branch, $n_-$ spins on the backward branch and $n_\beta$ spins on the imaginary-time branch of the contour ($n=n_++n_-+n_\beta$).
The weight of such a configuration is obtained by tracing over the dot and lead degrees of freedom and can be expressed in terms of two determinants of $n\times n$ matrices $N_\sigma^{-1}$:\cite{Gull08_ctaux} 
\begin{widetext}
\begin{eqnarray}
w(\{ (t_{1}, s_1),(t_{2}, s_2), \ldots (t_{n}, s_{n}) \})&=&(-i^{n_-})(i^{n_+})(K_t dt/2t)^{n_-+n_+}(K_\beta d\tau/2\beta)^{n_\beta}\prod_\sigma \det N_\sigma^{-1},\label{weight}\\
N_\sigma^{-1} &=& e^{S_\sigma}-(iG_{0,\sigma})(e^{S_\sigma}-I).
\end{eqnarray}
\end{widetext}
Here $(G_{0,\sigma})_{ij}=G_{0,\sigma}(t_i,t_j)$ is the $ij$ element of the $n\times n$ matrix of non-interacting Green functions 
\begin{equation}
G_{0,\sigma}(t,t')=-i\langle \text{T}_\mathcal{C} d_\sigma(t)d^\dagger_\sigma(t')\rangle_0
\label{eqn:G0input}
\end{equation}
computed using the possibly time-dependent chemical potentials and evaluated at the time arguments defined by the Ising spins. 
The quantity 
$e^{S_\sigma}=\text{diag}(e^{\gamma_1 s_1\sigma}, \ldots, e^{\gamma_n s_n \sigma})$ is a diagonal matrix depending on the spin variables (with $\gamma_i=\gamma_t$ for spins located on the real-time branches and $\gamma_i=\gamma_\beta$ for spins on the imaginary time branch). ${\text T}_\mathcal{C}$ is the contour ordering operator, 
which exchanges the 
product $A(t) B(t')$  of two operators if $t$ is earlier on the contour than $t'$ (a 
minus sign is added if the exchange involves an odd number of Fermi operators).

A Monte Carlo sampling of all possible spin configurations is then implemented based on the absolute value of the weights (\ref{weight}). 
The contribution of a specific configuration $c=\{ (t_{1}, s_1),(t_{2}, s_2), \ldots (t_{n}, s_{n}) \}$ to the current is given by \cite{Werner09} 
\begin{widetext}
\begin{eqnarray}
A_\sigma^c(t,t')&=&A_{0,\sigma}(t,t')+i\sum_{i,j=1}^n G_{0,\sigma}(t,t_{i})[(e^{S_\sigma}-I)N_\sigma]_{i,j}A_{0,\sigma}(t_{j}, t'),\label{eqA}
\label{tildeA}
\end{eqnarray}
\end{widetext}
with the first term on the right hand giving the contribution to the non-interacting current and the second term a correction due to the interactions. 
In Eq.~(\ref{tildeA})
\begin{equation}
A_{0,\sigma}(t,t')=\langle \text{T}_\mathcal{C} \tilde a^{L\dagger}_\sigma(t')d_\sigma(t)\rangle_0
\label{defA0}
\end{equation}
denotes a dot-lead correlation function of the noninteracting model for the composite left lead operator 
$ \tilde a^{L}_{\sigma}=\sum_p V^L_p a^L_{p,\sigma}$. The current expectation value is 
\begin{equation}
I(t)=-2\text{Im} \sum_\sigma[\langle A^c_\sigma(t,t) \phi_c\rangle/\langle \phi_c\rangle],
\end{equation}
where $\langle . \rangle$ denotes the Monte Carlo average and $\phi_c$ the phase of the weight of the configuration $c$.

In an interaction quench, the imaginary-time evolution is not explicitly considered in the Monte Carlo simulation and  temperature appears only as a parameter in the noninteracting Green functions (see Fig.~\ref{Lcontour}). Moreover, the latter depend only on time differences, and thus can be easily expressed in terms of their Fourier transform. 
Assuming a large band cutoff and neglecting the real part of the lead self-energy we find\cite{Jauho94, Werner09} 
\begin{widetext}
\begin{eqnarray}
G_0(t,t')&=&2i\sum_{\alpha=L,R}\int \frac{d\omega}{2\pi}e^{-i\omega(t-t')}\frac{\Gamma^\alpha(\omega)(f(\omega-\mu_\alpha)-\Theta_\mathcal{C}(t,t'))}{(\omega-\epsilon_d-U/2)^2+\Gamma^2},\label{G0}\\
A_0(t,t')&=&-2i\int \frac{d\omega}{2\pi}e^{-i\omega(t-t')}\frac{\Gamma_L(\omega) \Gamma_R(\omega) (f(\omega-\mu_L)-f(\omega-\mu_R))}{(\omega - \epsilon_d-U/2)^2+\Gamma(\omega)^2}\nonumber\\
&&+2\int \frac{d\omega}{2\pi}e^{-i\omega(t-t')}\frac{\Gamma_L(\omega)(\omega - \epsilon_d-U/2)(f(\omega-\mu_L)-\Theta_\mathcal{C}(t,t'))}{(\omega - \epsilon_d-U/2)^2+\Gamma^2}. 
\label{A0}
\end{eqnarray}
\end{widetext}

In the voltage quench, on the other hand, the interaction is non-vanishing on the imaginary time portion of the contour (Fig.~\ref{Lcontour}), while the chemical potential difference jumps instantaneously from zero (on the imaginary branch) to $V$ (on the real branches). Because of the time dependence of the chemical potentials, the noninteracting Green functions are not time translation invariant and we cannot express $G_{0,\sigma}$ and the dot-lead  correlator $A_{0,\sigma}$ in the form of a Fourier transform. 
Instead, those functions must be computed numerically from their equations of motion, as explained in the appendix.

\subsection{Optimization of the Monte Carlo sampling}

The sign (phase) problem in the weak-coupling CTQMC method grows exponentially with the average perturbation order on the real-time branches, which in turn is proportional to the simulation time, while operators on the imaginary time branch do not add significantly to the sign problem.  To reach long times or strong interactions, it is therefore important to reduce the average perturbation order on the real-time branches as much as possible. An essential point to note in this context is that in the particle-hole symmetric case, the parameters $K_x$ of the algorithm can be chosen such that only even perturbation orders appear in the expansion. In fact, for 
\begin{equation}
K_x=-xU/4
\end{equation}
the spin degree of freedom effectively disappears ($e^{\gamma s \sigma}=-1$) and the algorithm becomes the real-time version of Rubtsov's weak-coupling method\cite{Rubtsov05} for the particle-hole symmetric interaction term $H_U-K_x/x=U(n_{d,\uparrow}-\frac{1}{2})(n_{d,\downarrow}-\frac{1}{2})$. (For a detailed discussion of the equivalence between the Rubtsov and CTAUX methods for the Anderson impurity model, consult Ref.~\onlinecite{Karlis09}). The odd perturbation orders are continuously suppressed as $K_x$ approaches $-xU/4$. 
For $K_x=-xU/4+\delta$ and sufficiently small $\delta$, the average perturbation order can be reduced by about half compared to the $|K|=0.1$ used in the simulations presented in Ref.~\onlinecite{Werner09}. This in turn allows us to reach times and interaction strengths which are a factor of two larger. We note in passing that the suppression of odd perturbation orders was also essential in the nonequilibrium dynamical mean field calculations of Ref.~\onlinecite{Eckstein09}.

We next discuss some tricks to improve the efficiency of the current measurement. First, we rewrite Eq.~(\ref{eqA}) as
\begin{widetext}
\begin{equation}
A_\sigma^c(t,t')=A_{0,\sigma}(t,t')+\int ds_1 \int ds_2 G_{0,\sigma}(t,s_1) \Big\langle i \sum_{i,j=1}^n \delta_\mathcal{C}(s_1,t_{i})[(e^{S_\sigma}-I)N_\sigma]_{i,j}\delta_\mathcal{C}(s_2,t_{j})\Big\rangle A_{0,\sigma}(s_2, t'),
\label{X}
\end{equation}
\end{widetext}
where the variables $s_1$ and $s_2$ run over the entire contour and the contour delta function is defined by $\int ds \delta_\mathcal{C}(t,s)f(s)=f(t)$. It is therefore sufficient to accumulate the quantity
\begin{equation}
X_\sigma(s_1, s_2)=\Big\langle i \sum_{i,j=1}^n \delta_\mathcal{C}(s_1,t_{i})[(e^{S_\sigma}-1)N_\sigma]_{i,j}\delta_\mathcal{C}(s_2,t_{j})\Big\rangle. 
\end{equation}
Furthermore, it follows from Eq.~(\ref{weight}) that the weight of a Monte Carlo configuration changes sign if the last spin (corresponding to the largest time argument) is shifted from the forward contour to the backward contour or vice versa. Since the absolute value of the weight does not change, these two configurations will be generated with equal probability. As a result, all the terms in Eq.~(\ref{X}) which do not involve the last operator on the contour will cancel. It is therefore more efficient and accurate to accumulate 
\begin{align}
&X_\sigma(s_1, s_2)=\Big\langle
i(1-\delta(\{t_i\}))\sum_{i,j=1}^n x(s_1, i; s_2, j)\nonumber\\
&\hspace{5mm}+i\delta(\{t_i\})\sum_{l \text{ not last}}^n [ x(s_1,
\text{last}; s_2, l)+x(s_1, l; s_2, \text{last})] \Big\rangle,
\end{align}
with $x(s_1, i; s_2, j) \equiv \delta_\mathcal{C}(s_1,t_i)[(e^{\Gamma_\sigma}-1)N_\sigma]_{i,j}\delta_\mathcal{C}(s_2,t)$ and $\delta(\{t_i\})=1$ if $\max_i \text{Re}(t_i)>0$ and 0 otherwise.

Also, by comparing the contributions to the current of the original configuration and the one with the last operator shifted from the upper to the lower contour (or vice versa), one finds that they almost (but not completely) cancel. The errorbars on the current can thus be substantially reduced by appropriate symmetrizations of $X(s_1, s_2)$.

\section{Results: Interaction quench \label{interactionquench}}

\subsection{Convergence to the long time limit: large bias voltage}

Calculations based on an interaction quench from $U=0$ are particularly simple, because there are no interaction vertices (or spins) on the imaginary time branch, and only the real-time branches of the contour need to be considered in the simulation. Temperature enters only as a parameter in the lead correlators, making  it possible to treat arbitrary temperatures, including $T=0$.
%
\begin{figure}[t]
\begin{center}
\includegraphics[angle=-90, width=0.9\columnwidth]{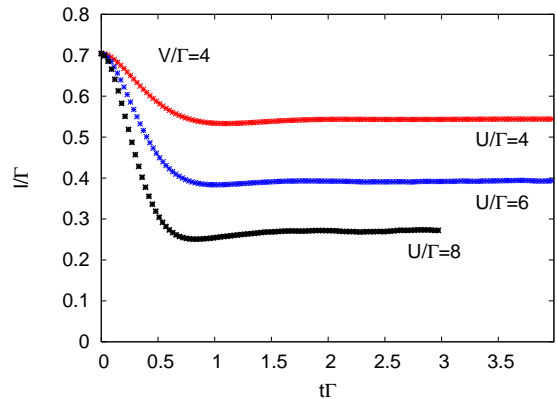}
\caption{Time evolution of the current for $V/\Gamma=4$ and different interaction strengths ($T=0$). In the initial state, the current is given by the steady state current through the non-interacting dot. At time $t=0$, the interaction is turned on. 
After a time of a few inverse $\Gamma$, the current saturates at the value corresponding to the steady state current in the interacting dot.
}
\label{i_t_v4}
\end{center}
\end{figure}
\begin{figure}[t]
\begin{center}
\includegraphics[angle=-90, width=0.9\columnwidth]{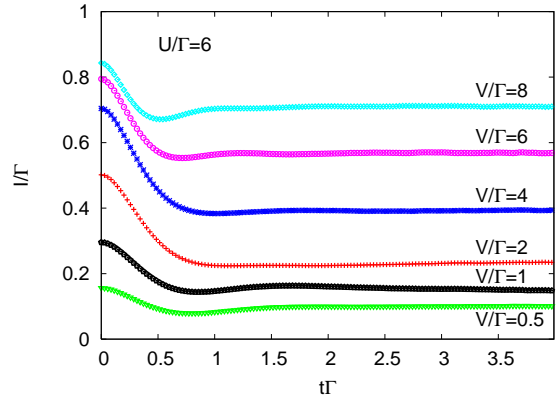}
\caption{Time evolution of the current for different voltage biasses and interaction strength $U/\Gamma=6$ ($T=0$). In the initial state, the current is given by the steady state current through the non-interacting dot. At time $t=0$, the interaction is turned on. 
}
\label{i_t_u6}
\end{center}
\end{figure}
At time $t=0$, the system is non-interacting but subject to an applied bias $V$, so a current $I_0(V)$ appropriate to the non-interacting model is flowing through the dot. At $t=0_+$ the interaction is turned on and the system relaxes into the steady-state configuration appropriate to the interacting model. Figure~\ref{i_t_v4} shows the time dependence of the current calculated for the large bias voltage $V/\Gamma=4$ and several interaction strengths. We see that the  transient behavior is such that  the current initially decreases sharply, overshoots and eventually relaxes more slowly back up  into the new steady state. The interaction-dependence of the steady-state current  is a consequence of the Coulomb blockade physics, apparent even at the large voltages studied here.   

For intermediate and large voltage bias ($V/\Gamma\gtrsim 2$) and not too large interaction ($U/\Gamma \lesssim 8$) the time required for convergence to the steady state is $t\Gamma\approx 2$, essentially independent of interaction strength. Given the scaling of the perturbation order (and hence the sign problem) with $U$ and $t$, interactions up to $U/\Gamma \lesssim 10$ are accessible with the current implementation.  A comparison of Fig.~\ref{i_t_v4} to Fig.~13 of Ref.~\onlinecite{Werner09} shows that the technical improvements introduced in this paper have substantially extended the range of applicability of the weak-coupling Monte Carlo method (about a factor 2-3 in $U$ or $t$) and allow us to obtain accurate results in the intermediate-to-strong correlation regime. 

In Fig.~\ref{i_t_u6} we plot the time evolution of the current for fixed $U/\Gamma=6$ and several voltage biasses. 
For voltages $V/\Gamma \gtrsim 2$, even though the transient behavior is clearly voltage-dependent, the current settles into the new steady state after a time $t\Gamma \approx 2$. However, as the voltage is decreased below $V/\Gamma\approx 2$ the transient time increases. At $V=\Gamma$ the long time limit is attained only for $t\Gamma \gtrsim 3$ and as $V$ is further decreased the approach to the asymptotic behavior becomes even slower.

\subsection{Convergence to the long time limit: small bias voltage}

\begin{figure}[t]
\begin{center}
\includegraphics[angle=-90, width=0.9\columnwidth]{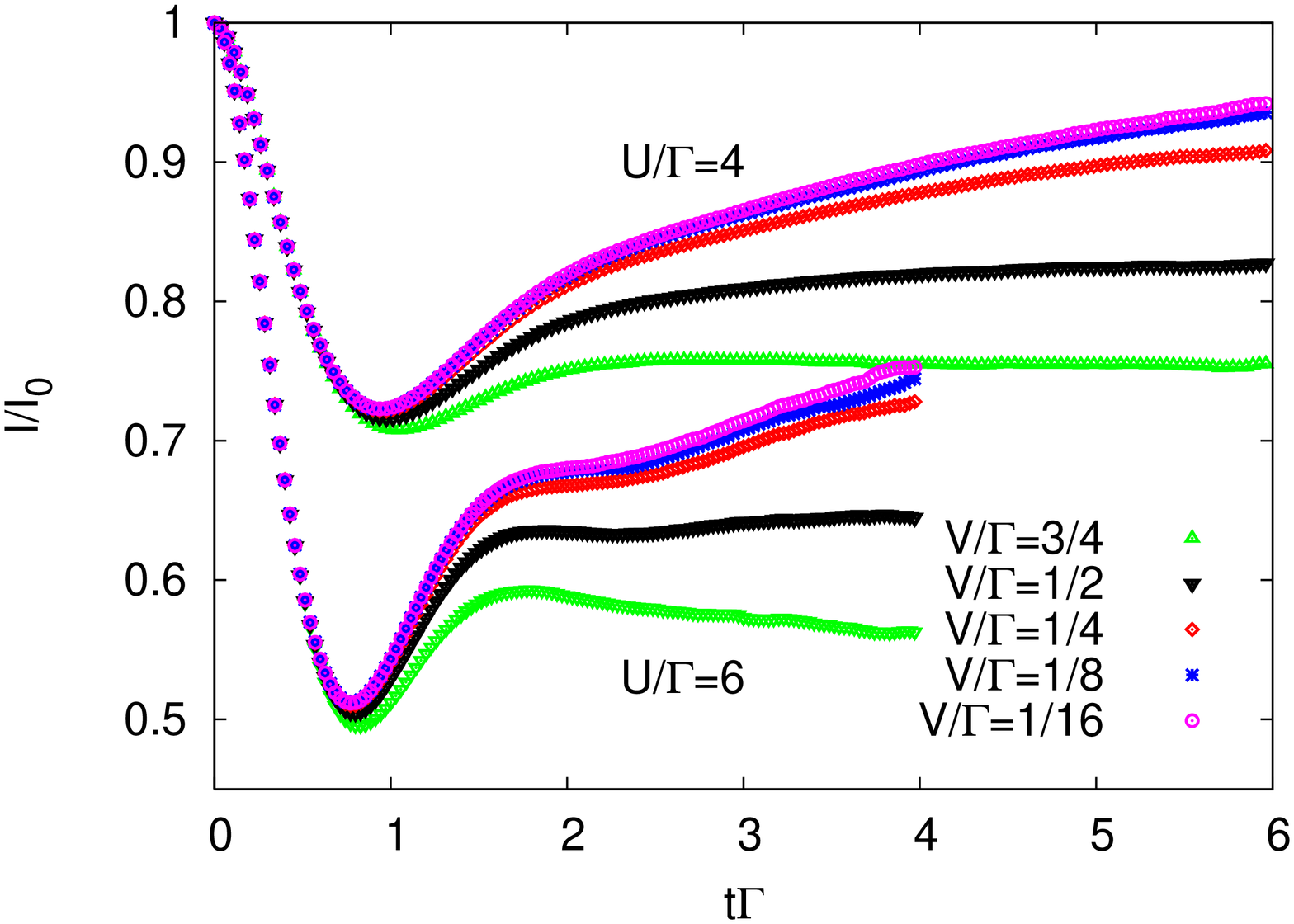}
\includegraphics[angle=-90, width=0.9\columnwidth]{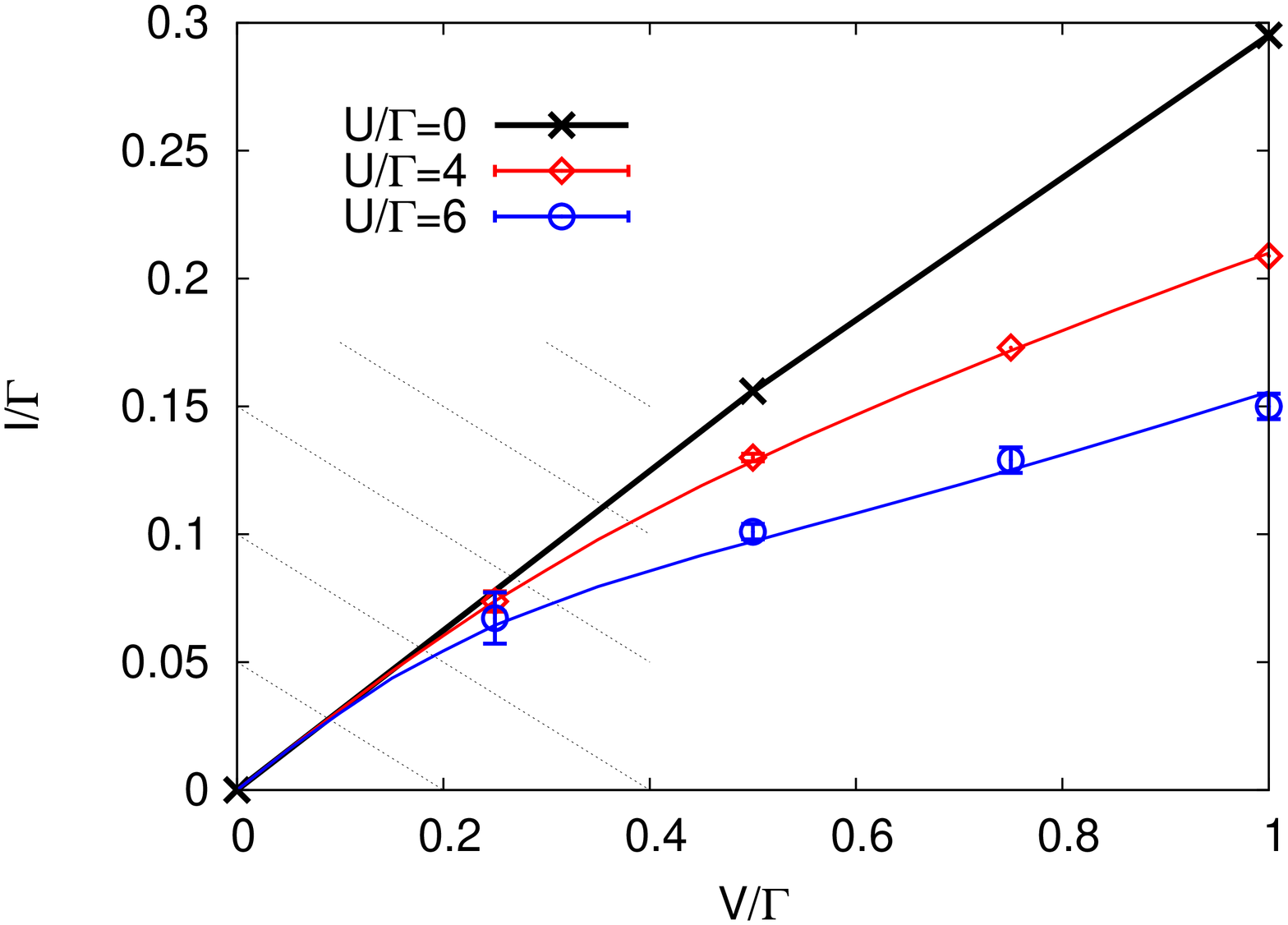}
\caption{Interaction quench in the small-voltage regime ($T=0$): The top panel shows the ratio of interacting to noninteracting current for $U/\Gamma=4$ and $U/\Gamma=6$ and indicated voltage biasses.
For $V/\Gamma \lesssim 0.5$ the time needed to reach the steady state grows much beyond the largest time accessible in the Monte Carlo simulation. 
Bottom panel: comparison of $U$-quench estimates (symbols, red and blue online) for the steady-state current to the noninteracting current (thick black line) and fourth order perturbation theory\cite{Fujii03}  (light lines, red and blue online). For $V/\Gamma \ge 0.5$ the Monte Carlo data show the value of $I(t_\text{max})$ with errorbar $\max_{t\in [t_\text{max}/2,t_\text{max}]}|I(t)-I(t_\text{max})|$ ($t_\text{max}\Gamma = 6$ for $U/\Gamma=4$ and $t_\text{max}\Gamma = 4$ for $U/\Gamma=6$). For $V/\Gamma=0.25$, we use $I \approx (I(t_\text{max})+I_0)/2$ with errorbar of size $(I_0-I(t_\text{max}))/2$.
}
\label{i_t_smallv}
\end{center}
\end{figure}

To better analyse the approach to steady state at small voltages we present in the upper panel of Fig.~\ref{i_t_smallv}  the time dependence of the current for several  smaller voltages and two interaction strengths. For better comparison, we plot here the ratio $I/I_0$ of the interacting current $I$ to the noninteracting current $I_0$. 
One sees that as $V$ is decreased or $U$ is increased the evolution of the current from the post-quench minimum to the long-time steady state value takes an increasingly long time. Since the longest accessible time is $t\Gamma\approx 6$ for $U/\Gamma=4$ and $t\Gamma\approx 4$ for $U/\Gamma=6$, the accurate measurement of $I$ becomes impossible in the small voltage regime. 
However the short-time transient behavior is  accessible at all voltages.  While the ratio $(I/I_0)(t)$ is clearly voltage dependent at higher biasses,  the data seem to converge as $V$ is reduced  to a non-trivial curve with a pronounced minimum  near an only weakly $U$-dependent time $t\Gamma\approx 1$. 

We believe that the increasingly slow convergence as $V\rightarrow 0$ is a signature of the Kondo effect, which is characterized by an energy scale which becomes exponentially small as $U$ increases. After the interaction quench, the Kondo resonance has to be built up as time progresses, and in the limit $V\rightarrow 0$, $T\rightarrow 0$ this requires an increasingly large number of interaction vertices and hence an increasingly long simulation time. On physical grounds one expects that the time needed to evolve into steady state is proportional to the inverse of the associated energy scale. 

Empirically, we find that the slow relaxation becomes an issue in the linear response regime, where the non-interacting and interacting currents are very similar. For $V/\Gamma\gtrsim 0.5$, where the interacting current is substantially smaller than $I_0$, a useful estimate of $I$ seems possible, even though in the voltage window up to $V/\Gamma\approx 2$ a small drift in the current may remain up to the longest accessible times. This drift makes it difficult to define reliable error bars on $I$, but it appears unlikely that the steady state value will differ from $I(t_\text{max})$ by more than the largest deviation in the window $[t_\text{max}/2, t_\text{max}]$, which we use as error estimate.
For $V/\Gamma=0.25$, an accurate estimate is not possible from the interaction quench procedure, but the current is very close to the linear response value, so the uncertainty is in fact not that important. 
This is illustrated in the bottom panel of Fig.~\ref{i_t_smallv}, which compares the Monte Carlo data to the noninteracting current and results from fourth order perturbation theory.\cite{Fujii03} The plot also indicates as hashed region the voltage range $V/\Gamma\lesssim 0.4$ where accurate measurements of the long-time limit become prohibitively difficult at $T=0$. We will see below that this roughly corresponds to voltages smaller than the Kondo temperature $T_K$.

\subsection{Temperature Dependence}

\begin{figure}[t]
\begin{center}
\includegraphics[angle=-90, width=0.9\columnwidth]{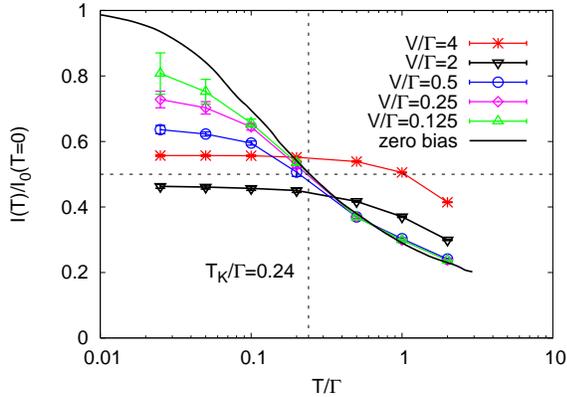}
\caption{Temperature dependence of the ratio of interacting current at temperature $T$ to noninteracting current at temperature zero for indicated values of the voltage bias. The interaction strength is $U/\Gamma=6$. 
The symbols show Monte Carlo results, the black line the analytical curve for $V\rightarrow 0$ extracted from Ref.~\onlinecite{Konik01} and plotted for $T_K/\Gamma=0.24$.
}
\label{i_beta}
\end{center}
\end{figure}

It is also of interest to examine the temperature dependence of the current. The interplay between voltage and temperature as the Kondo regime is approached presents an interesting problem. One expects that as the temperature is increased, the Kondo effect gets washed out and the simulations would therefore more readily converge even at small bias voltages. The temperature dependence of the current calculated from the interaction quench for $U/\Gamma=6$ and several values of the voltage bias is plotted in Fig.~\ref{i_beta}. In the linear response regime ($V/\Gamma=0.125$, $0.25$) the ratio of the interacting current $I(T)$ to the noninteracting current $I_0(T=0)$ exhibits a strong temperature dependence, even at $T/V\ll 1$. The temperature dependence arises because lowering the temperature strengthens the Kondo resonance and leads to an increase in the interacting current.  The temperature dependence for small voltage bias ($V/\Gamma=0.125$) approaches the analytical result for the temperature dependent zero-bias conductance in Ref.~\onlinecite{Konik01} and thus allows us to estimate (from the temperature at which $I(V\rightarrow 0)=I_0/2$) the Kondo temperature as $T_K/\Gamma\approx 0.24$, in good agreement with the {\it a priori} estimate from the standard formula\cite{Hewson}
\begin{equation}
T_K \approx U\Big(\frac{\Gamma}{2U}\Big)^{1/2}e^{-\pi U / 8\Gamma + \pi \Gamma / 2 U}.
\end{equation}
This formula is valid in the strong correlation regime and for $U/\Gamma=6$ yields $T_K/\Gamma=0.21$.

As $V$ is increased the temperature dependence is weakened. 
At intermediate values of $V$, in the Coulomb blockade regime ($V/\Gamma=2$), the current  has little temperature dependence at low $T$. 
At large voltage bias ($V/\Gamma=4$), correlation effects are already weakened due to the voltage, as is evident from the increase in $I/I_0$, and the almost perfect agreement with fourth order perturbation theory discussed in Section~\ref{sec:iv}. The current in this regime remains insensitive to temperature at low $T$.

\section{Results: Voltage quench \label{voltagequench}}
\label{vquench}

\subsection{Cutoff dependence}

An alternative procedure to calculate the steady state current of interacting quantum dots is to start from an interacting state in equilibrium ($V=0$) and turn on the voltage at $t=0_+$. While this approach is computationally more expensive and is restricted to nonzero temperatures, because it involves operators on the imaginary-time branch, it has the advantage that the Kondo resonance in the many-body density of states is present already in the initial state. The Kondo resonance is built up during the evolution along the imaginary-time branch, which does not add significantly to the sign problem. One might expect that this $V$-quench is particularly suitable to study the small voltage regime, because turning on a small voltage will not change the spectral function dramatically. 

Since the voltage quench has not yet been discussed in the previous literature, we will now analyze the properties of the current after such a $V$-quench in some detail, and in particular its dependence on the band width ($\omega_c$) and smoothness of the cutoff ($\nu$).
\begin{figure}[t]
\begin{center}
\includegraphics[angle=-90, width=0.9\columnwidth]{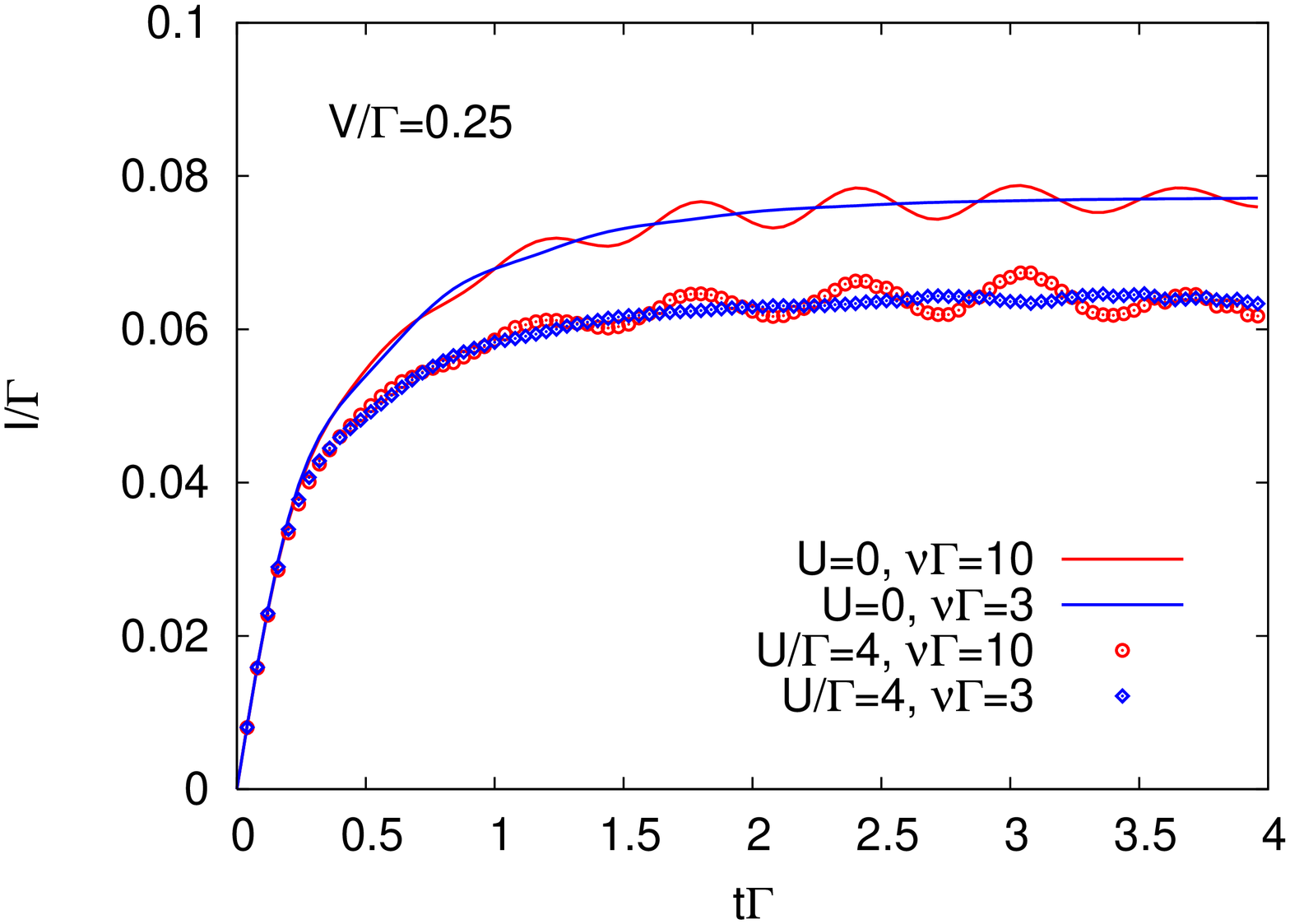}
\includegraphics[angle=-90, width=0.9\columnwidth]{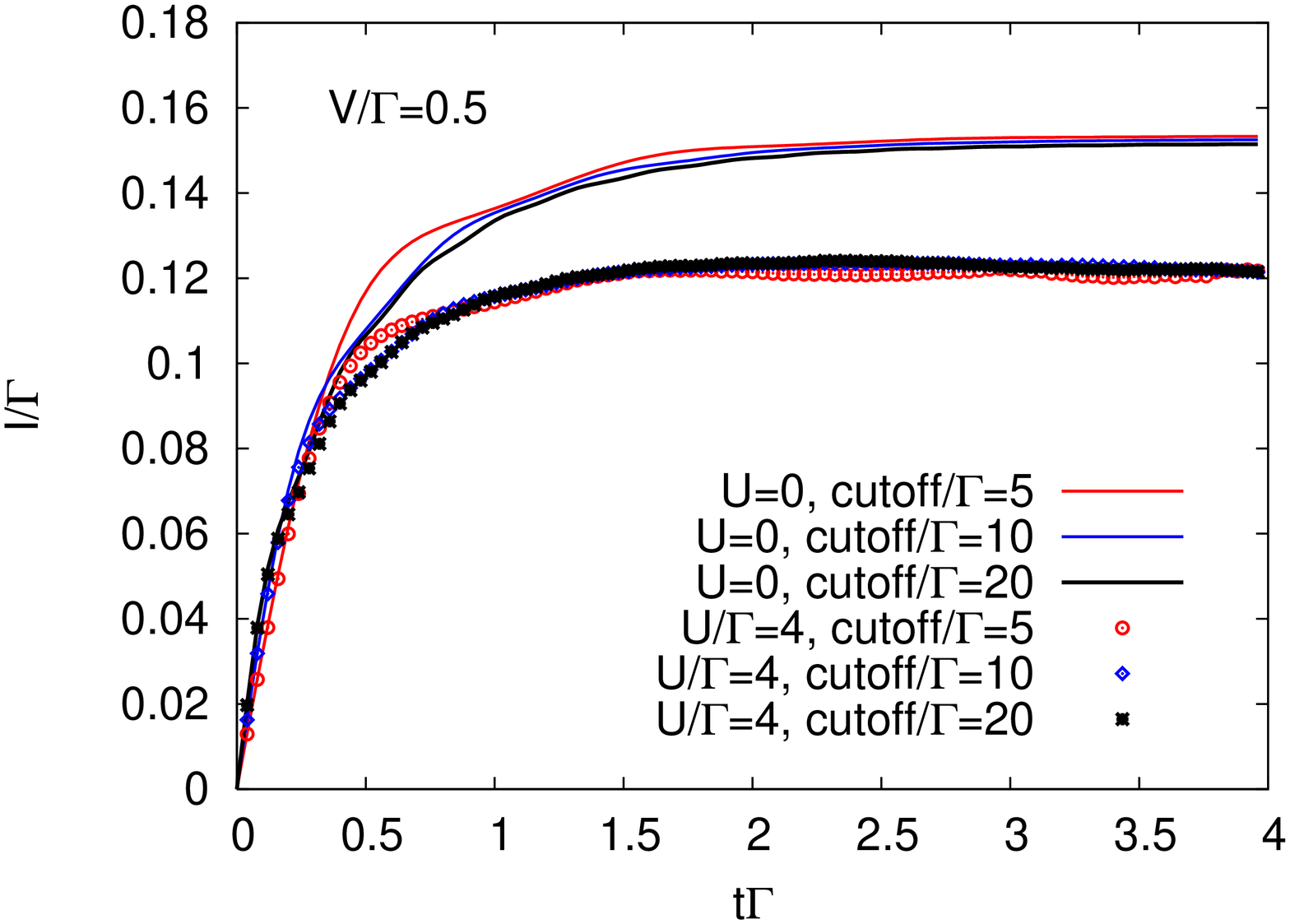}
\caption{Effect of the smoothing-parameter $\nu$ and cutoff $\omega_c$ for $\beta \Gamma=10$. The top panel shows the interacting ($U/\Gamma=4$) and noninteracting current for a quench to $V/\Gamma=0.25$. The red line and red circles correspond to $\nu\Gamma=10$ and cutoff $\omega_c/\Gamma=10$, the blue lines and blue diamonds to $\nu\Gamma=3$ and cutoff $\omega_c/\Gamma=10$. A sharp band edge (large value of $\nu$) leads to oscillations in the current which make it difficult to estimate the steady state value.
The bottom panel plots non-interacting and interacting currents for $V/\Gamma=0.5$, $\nu\Gamma=3$, $\beta \Gamma=10$ and indicated values of the cutoff. While the short-time behavior of the current is cutoff dependent, the steady state value is essentially cutoff-independent, as long as $\omega_c\gtrsim V$.
}
\label{nu_cut}
\end{center}
\end{figure}
The top panel of Fig.~\ref{nu_cut} shows the time evolution of the current in a model with $U/\Gamma=0$ (lines) and $U/\Gamma=4$ (symbols) if the voltage is suddenly increased to $V/\Gamma=0.25\ll \omega_c/\Gamma=10$. In the initial state, the system is in equilibrium, with no current flowing through the dot. After the voltage bias is turned on, the current increases. In the model with hard band cutoff ($\nu\Gamma=10$, red online) oscillations in the current appear which make it difficult to estimate the steady state value. A smoother band cutoff ($\nu\Gamma=3$, blue online) almost completely eliminates these oscillations. We will thus in the rest of this subsection show results for the ``smoothing parameter" $\nu\Gamma=3$.

The lower panel of Fig.~\ref{nu_cut} illustrates the dependence of the current on the cutoff value $\omega_c$. While the short time behavior of the current depends strongly on the bandwidth, the steady state value shows little cutoff dependence as long as $\omega_c$ is substantially larger than the applied voltage bias. This is consistent with the observation for the $U$-quench in Ref.~\onlinecite{Werner09}, where it was found that $\omega_c/\Gamma=10$ was enough to get accurate results up to $V/\Gamma=10$. Hence, we will choose $\omega_c/\Gamma=10$ for the rest of this paper.

\begin{figure}[t]
\begin{center}
\includegraphics[angle=-90, width=0.9\columnwidth]{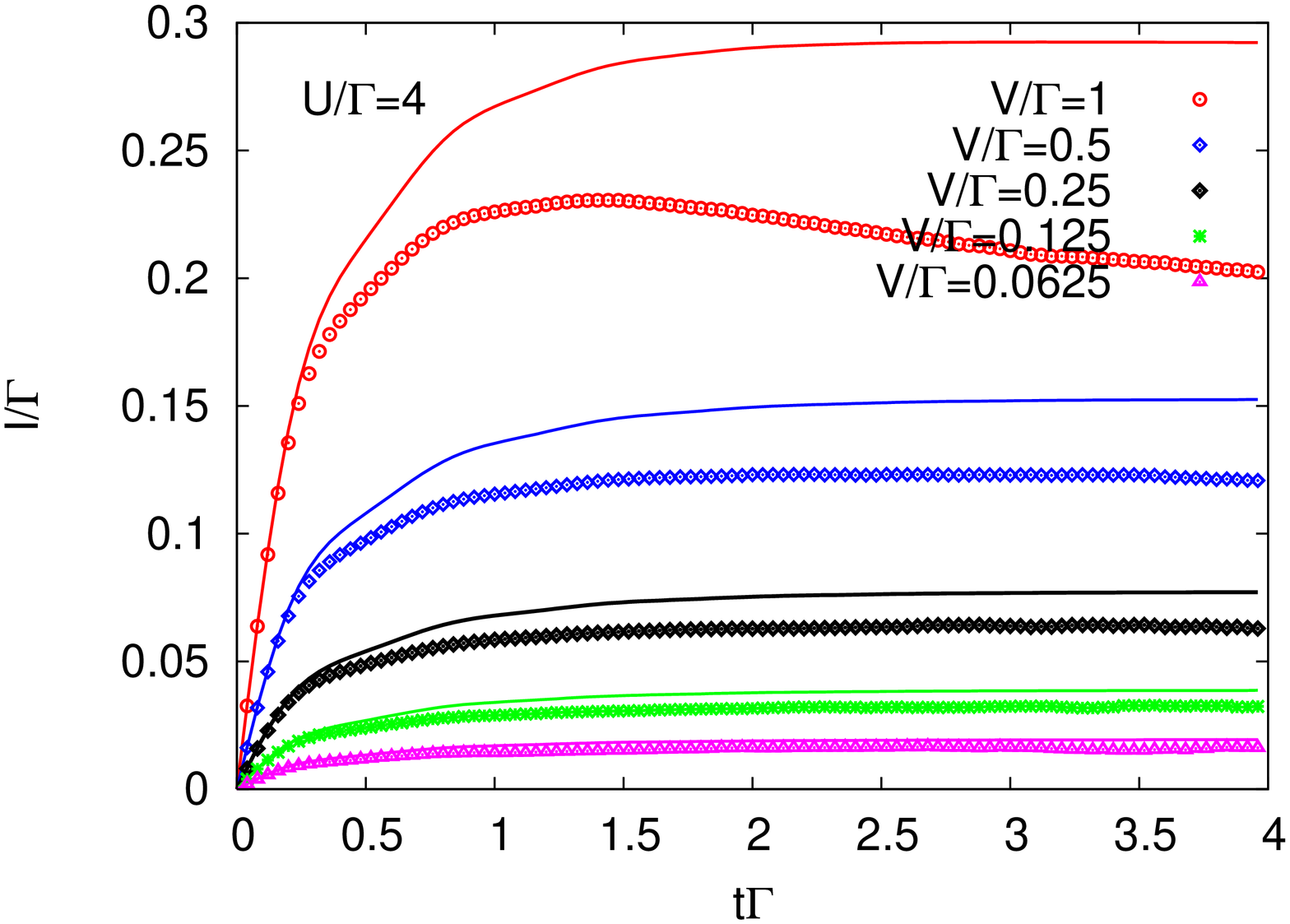}
\includegraphics[angle=-90, width=0.9\columnwidth]{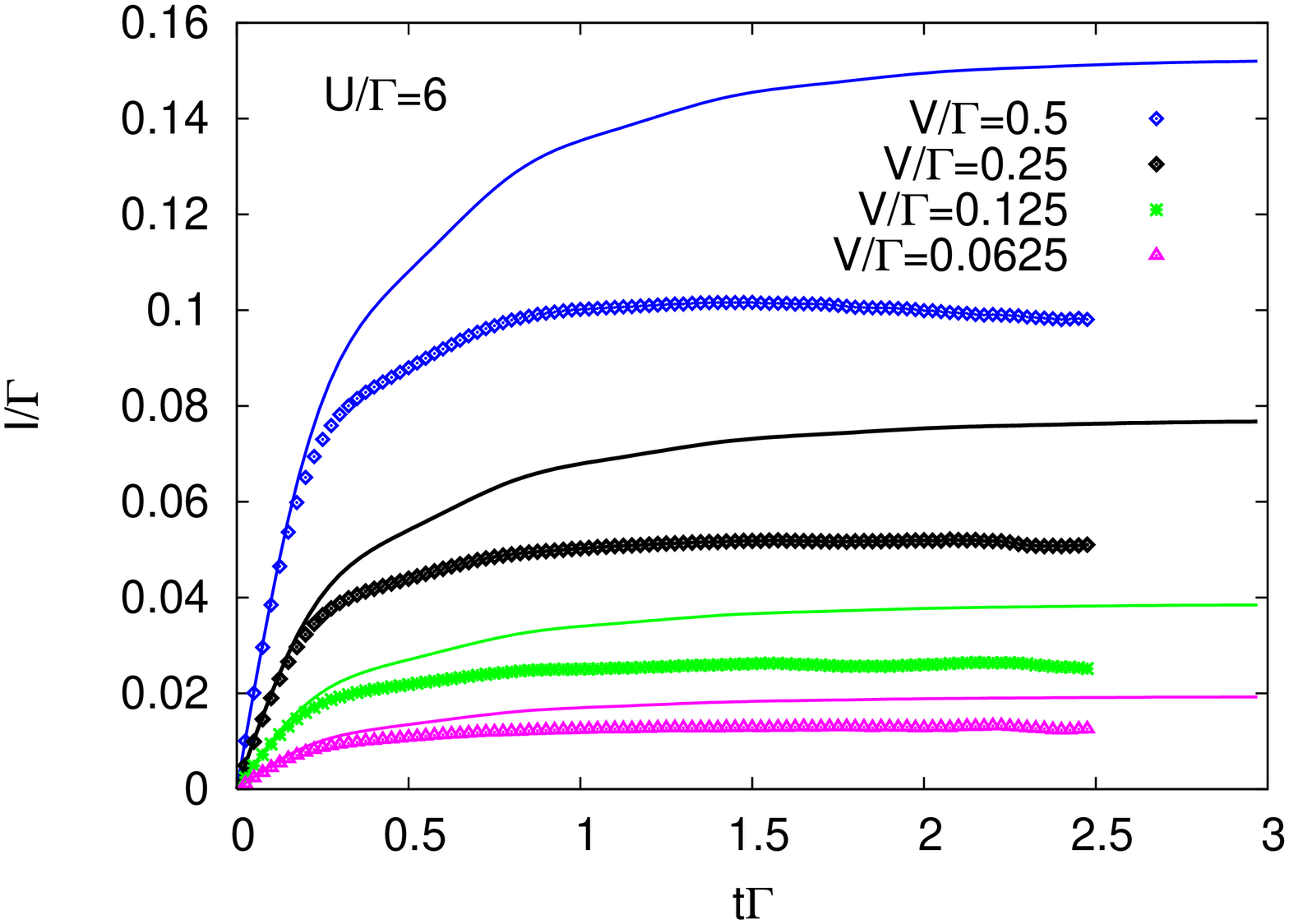}
\caption{Voltage dependence of the current for $U/\Gamma=4$ (top panel) and ($U/\Gamma=6$) bottom panel. Solid lined show the non-interacting current and symbols the interacting current. As the voltage reaches $V/\Gamma=1$ ($U/\Gamma=4$) or $V/\Gamma=0.5$ ($U/\Gamma=6$), the interacting current overshoots, resulting in a slow convergence to the steady state. For smaller voltages, however, the steady state current can be computed on the basis of $V$-quenches. All results are for $\beta\Gamma=10$, $\omega_c/\Gamma$=10, and $\nu/\Gamma=3$.}
\label{vu46}
\end{center}
\end{figure}

\subsection{Voltage dependence}

In Fig.~\ref{vu46} we plot the non-interacting and interacting currents for $\beta\Gamma=10$ and several values of the voltage bias. The top panel is for $U/\Gamma=4$ and the bottom panel for $U/\Gamma=6$. In the small voltage regime ($V/\Gamma\lesssim 0.5$) the interacting current increases monotonically with time and eventually settles into a steady state within the accessible time window. The $V$-quench therefore allows us to measure accurate steady state currents for finite temperature in the small voltage regime. However, once the voltage becomes too big (see $V/\Gamma=1$ in the top panel of Fig.~\ref{vu46}), the interacting current overshoots and only slowly settles into the steady state, making it impossible to measure an accurate steady state value using this approach. However, as shown in the previous section, the simulation based on the $U$-quench provides accurate results once $V/\Gamma\gtrsim 0.5$. The two simulation methods are therefore complementary in the sense that the $V$-quench works best for small voltage bias ($V/\Gamma\lesssim 0.5$) and the $U$-quench at larger voltage bias ($V/\Gamma\gtrsim 0.5$).

\subsection{Temperature dependence}

\begin{figure}[t]
\begin{center}
\includegraphics[angle=-90, width=0.9\columnwidth]{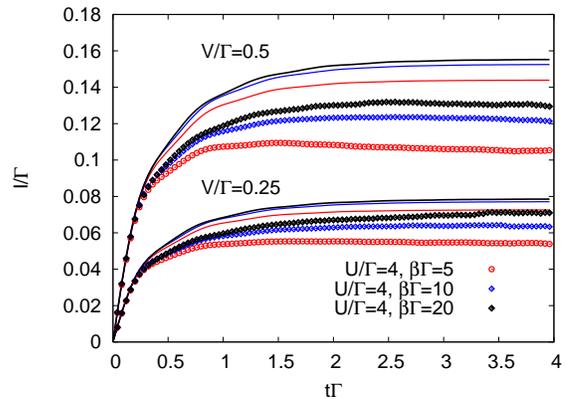}
\caption{Temperature effect on the current in the small voltage bias regime. Red circles, blue diamonds and black diamonds show the interacting current ($U/\Gamma=4$) for $V/\Gamma=0.25$ and $V/\Gamma=0.5$, while the red, blue and black curves indicate the corresponding non-interacting currents. As the temperature is lowered, the interacting current increases towards the non-interacting value.}
\label{beta}
\end{center}
\end{figure}

The temperature dependence of the current after a $V$-quench in the low-voltage regime ($V/\Gamma=0.25$ and 0.5) is shown in Fig.~\ref{beta}, which plots results for $U/\Gamma=0$ (lines) and $U/\Gamma=4$ (symbols) for $\beta\Gamma=5$, $10$ and $20$. A rather strong temperature dependence is evident, in particular in the interacting current. This is consistent with the $U$-quench data shown in Fig.~\ref{i_beta} and a consequence of the destruction of the Kondo resonance by temperature. Remarkably, a strong temperature dependence is observed even for $T\ll V$, which means that the applied voltage does not effectively raise the temperature to a value of order $V$. In this voltage regime the non-zero voltage state therefore is not simply equivalent to a thermal state. The nature of the correlations which give rise to the temperature dependence are an interesting subject for further investigation. 

\subsection{Comparison to the interaction quench}

\begin{figure}[t]
\begin{center}
\includegraphics[angle=-90, width=0.9\columnwidth]{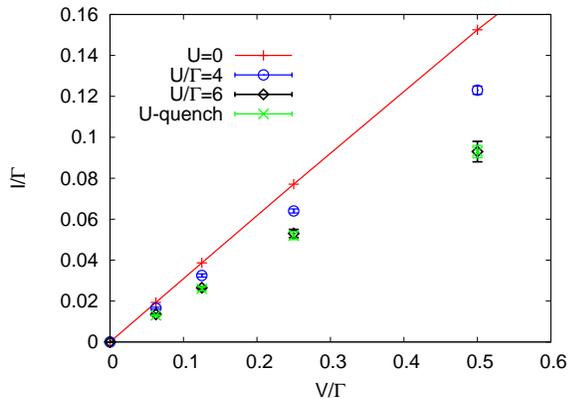}
\caption{
Current-voltage characteristics of the single-orbital Anderson impurity model in the small voltage regime for  $U/\Gamma=4$, 6 at $\beta\Gamma=10$. 
The blue circles and black diamonds show $V$-quench results. For comparison, we also plot finite temperature $U$-quench data ($U/\Gamma=6$, green crosses).
}
\label{currentsmall}
\end{center}
\end{figure}

In the $V$-quench calculations, for $U/\Gamma=6$, we can access temperatures down to $\beta \Gamma \approx 20$. At even lower temperatures, the perturbation order on the imaginary time branch becomes so large and the individual Monte Carlo updates so expensive that it is increasingly difficult to reach the very high statistical accuracy required for simulations with average signs of the order $10^{-3}$. Since the problem of slow convergence in $U$-quench calculations at small bias is considerably alleviated by finite temperature, it turns out that the accuracy of the latter approach matches that of $V$-quench calculations even at very small voltage bias (see $U/\Gamma=6$ data in Fig.~\ref{currentsmall}). For finite temperature simulations in the experimentally relevant temperature range ($\beta \Gamma \gtrsim 20$), the $U$-quench approach thus appears to be more powerful and sufficient to treat the entire voltage range. 
The good agreement between the $U$-quench and $V$-quench results in Fig.~\ref{currentsmall} furthermore shows that the steady state results obtained by the diagrammatic Monte Carlo method do not depend on the initial preparation of the system.

\section{$I$-$V$ characteristics \label{iv}}
\label{sec:iv}

We now apply the machinery described in the previous section to compute the current-voltage characteristics of the Anderson impurity model at half-filling. 
\begin{figure}[t]
\begin{center}
\includegraphics[angle=-90, width=0.9\columnwidth]{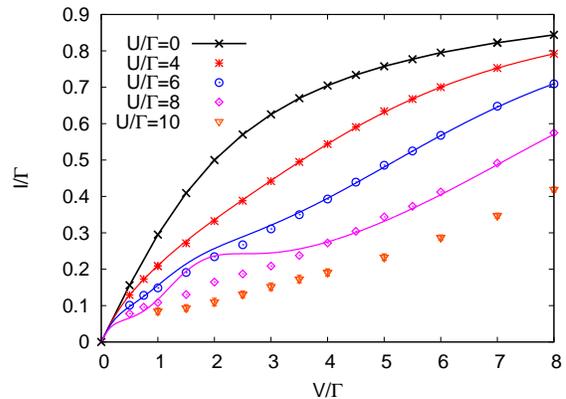}
\caption{Current-voltage characteristics of the single-orbital Anderson impurity model. The symbols show Monte Carlo data for $U/\Gamma=4$, 6, 8 and 10, while the lines correspond to the fourth order perturbation calculation of Ref.\onlinecite{Fujii03}. The Monte Carlo results have been obtained by means of $U$-quenches at $T=0$. Error bars are on the order of the symbol size.
}
\label{currentlarge}
\end{center}
\end{figure}
The initial rise of the current at finite temperature ($\beta\Gamma=10$) is shown in Fig.~\ref{currentsmall}. The blue circles and black diamonds have been obtained using the $V$-quench. The current-voltage characteristics in the $V\rightarrow 0$  limit becomes linear, although the slopes of the interacting and non-interacting models are not identical. This is the temperature effect on the Kondo resonance (particularly pronounced for large $U$) which was discussed in the previous sections. As the temperature is lowered to zero, the initial slope of the current approaches that of the non-interacting model. 

Figure~\ref{currentlarge} shows the $T=0$ result obtained using interaction quenches ($\omega_c/\Gamma=\nu\Gamma=10$, essentially the wide band limit). The black curve shows the monotonic increase of the non-interacting current with increasing applied bias voltage. The red, blue and pink lines show the interacting current for $U/\Gamma=4$, $6$ and $8$ predicted by fourth order perturbation theory.\cite{Fujii03} Consistent with analytical arguments,\cite{Ng88, Glazman88} the interacting current initially rises with the same slope as the non-interacting current, and reaches the non-interacting value also in the large-voltage limit. At intermediate values of $V$ the effect of interactions is to suppress the current (Coulomb blockade). In fourth order perturbation theory, a hump appears in the $I$-$V$ curve around $V/\Gamma=2$ for $U/\Gamma=6$ and $8$. At even larger $U$ (clearly outside the range of applicability) fourth order perturbation theory will presumably lead to a negative differential conductance at intermediate $V$. The appearance of this hump is related to the splitting of the Kondo resonance as discussed in Ref.~\onlinecite{Fujii03}. The Monte Carlo results for $U/\Gamma=4$, $6$, $8$, and 10 are shown by the red stars, blue circles, pink diamonds, and orange triangles, respectively. Since these are $U$-quench results for $T=0$, only $V/\Gamma\gtrsim 0.5$ data are shown. In the large voltage regime ($V/\Gamma \gtrsim 4$) the numerical results agree with the prediction from fourth order perturbation theory. Apparently, the fast decay of the Green functions for large voltage bias simplifies the diagram structure such that fourth order in $\Sigma$ is sufficient at $V/\Gamma \gtrsim 4$.
At intermediate voltages, $1 \lesssim V/\Gamma \lesssim 3$, differences between the Monte Carlo data and fourth order perturbation theory appear. The essentially exact numerical data show no prominent hump feature near $V/\Gamma=2$, and hence no negative differential conductance in the intermediate to strong correlation regime. The hump, and the associated splitting of the Kondo resonance, must therefore be an artefact of fourth order perturbation theory. This is consistent with the conclusion reached in Ref.~\onlinecite{Werner09} on the basis of (less accurate) hybridization expansion results. The data in Fig.~\ref{currentlarge} show that fourth order perturbation theory yields correct results over the entire voltage range for $U/\Gamma<4$. For larger interactions, and in particular around $V/\Gamma\approx 2$ more complicated self energy diagrams become important.

%
%

\section{Conclusions \label{conclusions}}

We have discussed the implementation of the weak-coupling continuous-time Monte Carlo method on the L-shaped Keldysh contour and the application of this formalism to the study of transport through a quantum dot. Calculations based on interaction quenches from the current carrying state at $U=0$ can be restricted to the real-time contours and provide accurate steady state currents for $V/\Gamma\gtrsim 0.5$ and interaction strenghts $U/\Gamma\lesssim 10$, for arbitrary temperature and bandwidth. 
At finite temperature, convergence into the steady state is considerably faster, which allows access to the small voltage regime.  As an alternative method, we have introduced 
calculations based on voltage quenches, which start from the interacting equilibrium state and which can be used to calculate the steady state current in the small voltage regime ($V/\Gamma\lesssim 0.5$) at finite temperature. Since the sign problem turns out to be essentially independent of the number of operators on the imaginary time branch, temperatures of order $\beta\Gamma=10$ can easily be dealt with. The $V$-quench approach is however not more efficient than finite-temperature $U$-quench calculations.   

We have used the methods to accurately compute the current-voltage characteristics of the half-filled Anderson-impurity model in the intermediate-to-strong coupling regime. Comparison to fourth order perturbation theory showed that the latter fails at voltages around $V/\Gamma\approx 2$ for $U/\Gamma>4$, but becomes accurate for $V/\Gamma\gtrsim 4$. The splitting of the Kondo resonance predicted by low order perturbation theory is an artefact not present in the numerical data. The results presented in this paper show that diagrammatic Monte Carlo is one of the most powerful numerical tools for the study of non-equilibrium systems. The accuracy of the improved weak-coupling approach and its range of applicability rivals or surpasses other state-of-the-art numerical approaches such as time-dependent DMRG.\cite{Kirino08, Heinrich-Meissner09} For most practical purposes the numerical problem of calculating the steady-state current through a half-filled Anderson impurity model with symmetrically applied voltage can be considered as solved.

An interesting, and presumably straight-forward extension of our work would be the study of asymmetrically applied bias voltages.  One of the optimizations of the Monte Carlo algorithm -- the suppression of the odd perturbation orders -- is however specific to the particle-hole symmetric model. Away from particle hole symmetry, odd perturbation orders contribute to the current and therefore must be considered in the simulation. This leads to an increase in the average perturbation order and to a more severe sign problem, such that the accessible times and interaction strengths will be reduced. The optimal choice of the $K_x$-parameters in the particle-hole asymmetric case is an open problem for future investigations. Another issue which should be considered is the optimal shape of the $U$- or $V$-quench. By slowly ramping up the  interaction or voltage bias, it may be possible to avoid overshooting and thus observe a faster relaxation of the current into the steady state.

\acknowledgements

PW and ME are supported by the Swiss National Science Foundation (Grant PP002-118866), TO by a Grant-in-Aid for Young Scientists (B) from MEXT, and AJM by the US National Science Foundation Division of Materials Research under grant DMR-0705847. This work also benefitted from the academic guest program (Center for Theoretical Studies) of ETH Zurich (TO) and the hospitality of the Aspen Center for Physics (PW). We thank N.~Tsuji, T.~Fujii and K.~Ueda for helpful discussions. The simulations were performed on the Brutus cluster at ETH Zurich using a code based on ALPS.\cite{ALPS}

\begin{appendix}
\section{Noninteracting Green function for the voltage quench}

\newcommand{\tmin}{t_\text{min}}
\newcommand{\tmax}{t_\text{max}}
\newcommand{\expval}[1]{\langle#1\rangle}
\newcommand{\CC}{\mathcal{C}}
\newcommand{\TC}{\text{T}_{\CC}}
\newcommand{\intC}{\int_\CC}
\newcommand{\ret}{{\text{r}}}
\newcommand{\adv}{{\text{a}}}
\newcommand{\mat}{{\text{\tiny M}}}
\newcommand{\tv}{{\makebox{$\neg$}}}
\newcommand{\vt}{{\reflectbox{$\neg$}}}
\newcommand{\les}{<}
\newcommand{\lar}{>}

In this appendix we present the formalism needed for the voltage quench,
evaluating noninteracting Green functions on the L-shaped contour with
lead chemical potential $\mu_\alpha$ equal to the equilibrium value $\mu(0)$
for times on the imaginary contour and with arbitrary time-dependence $\mu_\alpha(t)$
on the real time portions of the contour. The results of the paper correspond to  
$\mu_\alpha(t)=\mu(0)\pm V/2$.

Because the voltage bias is time dependent, noninteracting Green functions cannot
be expressed in the form of a Fourier transform, and instead they are computed 
numerically by the solution of their equations of motions in real (imaginary) time.
A closed set of equations is obtained if one considers the noninteracting 
dot Green function [Eq.~(\ref{eqn:G0input})],
\begin{equation}
\label{g0-def}
G_{0,\sigma}(t,t')=
-i\expval{\TC d_{\sigma} (t) d_\sigma^\dagger(t')}_0,
\end{equation}
the hybridization of the dot to a single bath level
\begin{equation}
G_{p,\sigma}^\alpha(t,t')=
i\expval{\TC d_\sigma(t) a_{p,\sigma}^{\alpha\,\dagger} (t')}_0,
\end{equation}
and the dot-decoupled Green function of a single bath state,
\begin{equation}
\label{geps-def}
g_{p,\sigma}^\alpha(t,t')=
-i\expval{\TC \,a_{p,\sigma}^\alpha (t) \,a_{p,\sigma}^{\alpha\,\dagger}(t')}_{0,V_p^\alpha=0}.
\end{equation}
Here $t,t'$ are arbitrary points on the real or imaginary portions of the contour, 
the time evolution is performed with $U=0$ but time-dependent voltage bias, and 
$\expval{\cdot}_0$ is the grand-canonical expectation value in the noninteracting
initial state (at $\mu=\mu(0)$). The contour-ordering operator $\TC$ exchanges the 
product $A(t) B(t')$  of two operators if $t$ is earlier on the contour than $t'$ (a 
minus sign is added if the exchange involves an odd number of Fermi operators).
Equations of motions for the Green functions (\ref{g0-def}) to (\ref{geps-def})
are obtained from taking time-derivatives and evaluation of the resulting commutators,
\begin{align}
\label{app-eom-g0}
&[i\partial_{t'}+\epsilon_d] G_{0,\sigma}(t,t')
=
\sum_p G_{p,\sigma}^\alpha(t,t') V_p^\alpha
-
\delta_\CC(t,t'),
\\
\label{app-eom-F}
&[i\partial_{t'} + \epsilon_{p\sigma}^\alpha-\mu_\alpha(t')]
\,G_{p\sigma}^\alpha(t,t')
=
G_{0,\sigma}(t,t')\,(V_p^\alpha)^*,
\\
\label{app-eom-geps}
&[i\partial_{t'} + \epsilon_{p\sigma}^\alpha - \mu_\alpha(t')]
\,g_{p\sigma}^\alpha(t,t')
=
-\delta_\CC(t,t').
\end{align}
Note that when $t=-i\tau$ is on the imaginary branch of the contour, the time derivative
is given by $\partial_t=i\partial_\tau$. The contour delta function is given by 
$\delta_\CC(t,t')=\partial_t \Theta_\CC(t,t')$, where $\Theta_\CC(t,t')=1$ if $t$ is 
later on $\CC$ than $t'$ and zero otherwise. Eqs.~(\ref{app-eom-g0}) to (\ref{app-eom-geps}) 
have a unique solution, provided that the contour Green functions satisfy an antiperiodic boundary 
condition on the contour $\CC$ in both time arguments.

Equation (\ref{app-eom-geps}) can be solved explicitly,
\begin{align}
\label{gbath}
g_{p\sigma}^{\alpha}(t,t')
&\equiv 
g_\alpha(t,t';\epsilon_{p\sigma}^{\alpha}),
\\
\label{geps}
g_\alpha(t,t';\epsilon)
&=
-i\big[\Theta_\CC(t,t')-f(\epsilon-\mu(0))\big]
\nonumber\\
\times \exp\Big(
&i\int_0^{t'} \!\!d\bar t  
\,[\epsilon-\mu_\alpha(\bar t)]
-i\int_0^{t} \!\!d\bar t\, [\epsilon-\mu_\alpha(\bar t)]
\Big),
\end{align}
where $\int_0^t d\bar t$ is along the contour. Furthermore, one can show from 
Eqs.~(\ref{app-eom-F}) and (\ref{app-eom-geps}) that the solution of 
Eq.~(\ref{app-eom-F}) is given by
\begin{equation}
\label{F}
G_{p\sigma}^{\alpha}(t,t')
=
\int ds \,
G_{0,\sigma}(t,s)\,
(V_p^\alpha)^*\,
g_{p\sigma}^{\alpha}(s,t'),
\end{equation}
where the integral runs over the whole contour.
This expression is inserted into Eq.~(\ref{app-eom-g0}) in order to derive 
a single closed equation for $G_0$,
\begin{equation}
\label{app-eom-g0-2}
[i\partial_{t'}+\epsilon_d] G_{0,\sigma}(t,t')-
\int \!ds \, G_{0,\sigma}(t,s)\Delta_\sigma(s,t') 
=
-\delta_\CC(t,t').
\end{equation}
Here the sum over bath states has been condensed into the integral 
over the hybridization function (\ref{Gamdef})
\begin{align}
\label{deltadef}
\Delta_{\sigma}(t,t')
&=\sum_\alpha \Delta_{\sigma}^{\alpha}(t,t'),
\\
\label{deltaalpha}
\Delta_{\sigma}^{\alpha}(t,t')
&\equiv
\sum_p |V_p^\alpha|^2 g_{p\sigma}^\alpha(t,t')
\\
\label{intgamma}
&= \int \!d\epsilon\, \frac{1}{\pi}\Gamma^{\alpha}(\epsilon)\,
g_{\alpha}(t,t';\epsilon).
\end{align}
In practice, we determine $\Delta$ from Eqs.~(\ref{intgamma}), (\ref{geps}) and 
(\ref{Gamdef}). 

Equation~(\ref{app-eom-g0-2}) is an integrodifferential equation on the contour $\CC$.
Its solution is equivalent to a boundary value problem for the 
imaginary time component of the Green function and initial value problems for 
the components involving real time-arguments. The equation is solved numerically, 
using Langreth rules for the decoupling of real and imaginary time components  
[see Ref.~\onlinecite{keldyshintro}].

The correlator (\ref{defA0}) which enters Eq.~(\ref{eqA}) for the current is 
by definition given by
\begin{equation}
\label{a1}
A_{0,\sigma} (t,t')
=
i\sum_p G_{p\sigma}^L(t,t') V_{p\sigma}^L. 
\end{equation}
Using Eqs.~(\ref{F}) and (\ref{deltaalpha}), this function
can be obtained from the contour integral 
\begin{equation}
\label{afinal}
A_{0,\sigma}(t,t')
=
\int \!ds\, G_{0,\sigma}(t,s)\Delta_{\sigma}^L(s,t').
\end{equation}

Note that the equations of motion (\ref{app-eom-F}) and (\ref{app-eom-geps})
also hold in the interacting case, with $G_0$ replaced by $G$. Hence 
Eqs.~(\ref{a1}) and (\ref{afinal}) are still valid in the interacting case with
the same replacement, and the interacting current can be obtained 
directly from the interacting dot Green function. This procedure is however equivalent 
to the approach which is used in the present paper, where the Green 
function is not measured and the current is obtained instead from Eq.~(\ref{eqA}).

\end{appendix}

\end{document}